\documentclass[fleqn,usenatbib]{mnras}

\usepackage{newtxtext,newtxmath}

\usepackage[T1]{fontenc}
\usepackage{ae,aecompl}

\usepackage{graphicx}	
\usepackage{amsmath}	
\usepackage{amssymb}	
\usepackage{multirow}   

\title{Statistical Modeling of an astro-comb for high precision radial velocity observation}

\author[F. Zhao et al.]
{
Fei Zhao,$^{1}$
Gang Zhao,$^{1}$ 
Yujuan Liu,$^{1}$
Liang Wang,$^{4,5}$
Huijuan Wang,$^{1}$
Hongbin Li,$^{1}$
\newauthor 
\ Huiqi Ye,$^{2,3}$
Zhibo Hao,$^{2,3}$
Dong Xiao,$^{2,3}$
Junbo Zhang,$^{1}$
Hanna Kellermann$^{4,5}$
\newauthor
\ and Frank Grupp$^{4,5}$
\\
$^{1}$Key Laboratory of Optical Astronomy, National Astronomical Observatories, Chinese Academy of Sciences, Datun Road 20A, Beijing, \\  \ 100012, China\\
$^{2}$National Astronomical Observatories Nanjing Institute of Astronomical Optics \& Technology, Chinese Academy of Sciences, Nanjing, \\ \ Jiangsu 210042, China\\
$^{3}$Key Laboratory of Astronomical Optics \& Technology, Chinese Academy of Sciences, Nanjing, Jiangsu 210042, China \\
$^{4}$Max-Planck-Institut f\"{u}r extraterrestrische Physik, Giessenbachstrasse, D-85748 Garching, Germany \\
$^{5}$Universit\"{a}ts-Sternwarte M\"{u}nchen, Scheinerstr. 1, D-81679 M\"{u}nchen, Germany
}

\date{Accepted 2018 October 06. Received 2018 October 06; in original form 2018 August 31}

\pubyear{2018}

\begin{document}
\label{firstpage}
\pagerange{\pageref{firstpage}--\pageref{lastpage}}
\maketitle

\begin{abstract}
The advent of the laser frequency comb as the wavelength calibration unit allows us to measure the radial velocity at $cm\ s^{-1}$ precision level with high stability in long-term, which enable the possibility of the detection of Earth-twins around solar-like stars.
Recent study shows that the laser frequency comb can also be used to measure and study the precision of the instrumental system including the variations of line profile and the systematic uncertainty and instrumental drift.
In this paper, we present the stringent analysis of a laser frequency comb(LFC) system with 25GHz repetition frequency on a R$\sim$50,000 spectrograph with the wavelength spanning from 5085\AA \ to 7380\AA. 
We report a novel fitting model optimized for the comb line profile, the constrained double Gaussian. The constraint condition is set as $\left|\mu_{1,2} - \mu \right| <\sqrt{2ln2}\sigma$. We introduce Bayesian information criterion to test various models. Compared to the traditional Gaussian model, the CDG(Constrained Double Gaussians) model provides much better goodness of fit.
We apply the CDG model to the observed comb data to demonstrate the improvement of RV precision with CDG model. We find that  the improvement of CDG model is about 40\%$\sim$60\% for wavelength calibration precision.
We also consider the application to use the LFC and CDG model as a tool to characterize the line shape variation across the detector.
The motivation of this work is to measure and understand the details of the comb lines including their asymmetry and behaviors under various conditions, which plays a significant role in the simultaneous calibration process and cross-correlation function method to determine the Doppler shift at high precision level.

\end{abstract}




\section{Introduction}
\label{sec1}
The golden era for the detection of exoplanets is enabled by the radial velocity(RV) method with precise astronomical spectroscopy. Beginning from the discovery of substellar companions\citep{1988ApJ...331..902C, 1989Natur.339...38L} and the validation of the first extra-solar planet orbiting a main-sequence star\citep{1995Natur.378..355M}, the study of Doppler meansurements has become a vibrant field in exoplanets research\citep{2015ARA&A..53..409W}.
To date, more than 3800 exoplanets are confirmed in which nearly 750 were discovered by Doppler method\footnote[1]{https://exoplanet.eu/ \& https://exoplanetarchive.ipac.caltech.edu/}. Among them, there are about 12\% super Earth and about 17\% Mini-Neptune(by mass classification scheme)\citep{2007ApJ...656..545V, 2009Natur.462..891C}. On the road of finding terrestrial exoplanets in habitable zone or even Earth-twins, high precision Doppler measurements play a vital role. 
Combined with transit observations, they provide the planetary mass for the measurement of the bulk density and characterization of the planet's structure\citep{2017ApJ...839L...8M, 2017A&A...608A..35C, 2018AJ....155..257S}.
The amplitude of the RV for a solar mass star due to an Earth mass planet at 1 $AU$ is about 9.6 $cm \ s^{-1}$ (or 174 kHz shift in frequency at 5500\AA). For this planet at 0.05$AU$, this amplitude can increases to about 1 $m \ s^{-1}$. 
With the goal of achieving a precision better than 1 m/s, it is necessary to identify and overcome the limitation on Doppler measurements. 
To understand and characterize the RV uncertainties, we briefly divide them into two major categories depending on their origins: instrumental\citep{2002Msngr.110....9P} and stellar\citep{2005ApJ...635.1281K, 2013ApJ...770..133H, 2014ApJ...789..154D}.
The instrumental uncertainty could dominate the RV precision especially in $cm \ s^{-1}$ level.
The research on RV noise caused by stellar photospheres will become more effective if we have a better understanding of the noise characteristics from instruments.
There are several main factors from instrumental aspect that induce RV uncertainties, such as wavelength calibration, the stability of the light injection for optics, detector effects etc.\citep{2012Msngr.149....2L}.
Among these, the wavelength calibration is a crucial step especially for an instrumental system based on simultaneous calibration method and cross-correlation function(CCF) technique to determine Doppler shifts\citep{1996A&AS..119..373B, 2007A&A...468.1115L}.

Originally, the detector can only record the number of photons or the light's intensity as a function of pixel position without the information of wavelength. In order to determine the continuous wavelength distribution, we should convert the pixel space into the wavelength space by observing a calibration lamp as reference. One of these solutions is to use a Thorium-Argon hollow cathode lamp(ThAr lamp). With the knowledge of the standard wavelength of thorium emission lines measured in laboratory, we can fit a calibration curve (or wavelength solution) with high order polynomials. For each echelle order, this fitting function translates the pixel position information into the wavelength information. 
A typical precision level provided by ThAr lamps is $10^{-7}$ to $10^{-9}$(or a few $m \ s^{-1}$ to several tens of $cm \ s^{-1}$ scaled by the speed of light)\citep{2012Msngr.149....2L}.
In the case of ThAr lamp, the wavelength calibration precision is limited by the line blending, the irregular intensity and space density of thorium emission lines, the lamps age, the contamination of strong emission lines from adjacent orders etc. 
Another calibration method is the absorption cell. Iodine cell technique is widely used to calculate the Doppler shifts. With the narrow absorption $I_2$ lines, we can model the spectrum by taking into account the variations of instrumental profile(IP)\citep{1996PASP..108..500B}. The limitation of iodine cell method mainly includes the loss of the starlight via the absorption gas cell, the contamination of the spectral lines and the narrow wavelength coverage.A typical RV precision level obtained by iodine cell is about several $m \ s^{-1}$\citep{2000A&A...362..585E}.

However, such a calibration precision level is not sufficient for detecting the population of terrestrial exoplanets. To estimate how small the shift is, we consider an R=50,000 spectrograph, its resolution unit will be expressed as 0.11\AA \ at 5500\AA. Meanwhile, the radial velocity resolution unit is $\Delta v$ = c/R = 6000 $m \ s^{-1}$. Typically, $\Delta \lambda$ should cover 2 pixels on detector based on Nyquist sampling. Thus the velocity resolution is equal to 3000 $m \ s^{-1}$ per pixel. For a 0.1 $m \ s^{-1}$ Doppler shift on a 12$\mu$m pixel size CCD, the physical shift in pixels is $4 \times 10^{-10}$m which is about 7.5 times the radius of a Hydrogen atom.
For such a exquisite shift, it is a challenge for wavelength calibration methods to measure and calculate. 


Beyond the traditional wavelength calibration methods like iodine cells and Th-Ar lamps, laser frequency combs(LFC)\citep{1999OptCo.172...59R, 2002fqml.conf..253U} can offer a promising solution to the wavelength calibration with ultra high precision and long-term stability\citep{2007MNRAS.380..839M, 2008Natur.452..610L, 2008Sci...321.1335S}. An LFC, based on femtosecond pulse mode-locked laser, can produce thousands of uniformly spaced modes in frequency space as a function of $f_n=f_0+n \cdot f_{rep}$, where $f_n$ is the frequency of the nth mode. The $f_0$ is the carrier envelope offset frequency which provide an offset for each mode. Here n is typically a large integer approximately in the range of $10^{5} -10^{6}$. The frequency spacing between each adjacent mode is a constant(which equals to the repetition frequency $f_{rep}$). The frequencies are synchronized with the absolute ratio frequency reference such as an atomic clock or a GPS(global positioning system), which makes the astronomical laser frequency comb(astro-comb) meet the requirements of an ideal tools for high precision wavelength calibration.
In the work of \citet{2012MNRAS.422..761M}, the authors studied the variation of Intra-pixel sensitivity(IPS) by using a moded-locked fiber-based laser frequency comb with $f_{rep}$=90MHz and 1.03Ghz after filtered by a designed Fabry-Perot(FP) cavity. With the symmetrical fitting model, they found the change of averaged IPS deviates by less than 8\% level.
\citet{2015ApJ...814L..21D} reported the observations of the Sun as a star with an astro-comb as the simultaneous wavelength calibrator. This work showed several tens $cm \ s^{-1}$ RV precision during 7 days time scale.

Nowadays, the laser frequency comb becomes a mature technology applied in several astronomical spectrographs worldwide. One of the available turn-key LFC is from Menlo system\citep{2008Sci...321.1335S, 2014SPIE.9147E..1CP}. 
In recent years, the Menlo combs are used or in operation at several telescopes, such as HARPS at 3.6m telescope in La Silla, the VTT solar telescope in Tenerife, the FOCES spectrograph in Wendelstain observatory at USM and the HRS(High Resolution Spectrograph) on 2.16m telescope in Xinglong observatory in China. 
In \citet{2010MNRAS.405L..16W}, the authors showed the evidence of the CCD stitching pattern revealed by using the Menlo system's 18GHz LFC mounted on HARPS. There are discontinuities on each 512 pixel position for the wavelength solution curve due to the variations of the intra-pixel distance and sensitivity at the borders of the stitching. By comparing the signals in both fibers(channels) simultaneously, a remarkable short-term repeatability of 2.5cm/s can be achieved on HARPS\citep{2012Natur.485..611W}. Since February 2016, a 25GHz LFC was installed in HRS\citep{2001ChJAA...1..555Z, 2016PASP..128k5005F}, which is a fiber-fed spectrograph with R$\sim$50,000 at 0.19mm slit width. In this paper, we demonstrate the calibration results of this HRS-comb system and analysis the line profile variations based on an optimized fitting model. 

\begin{figure*}
	\includegraphics[width=\textwidth]{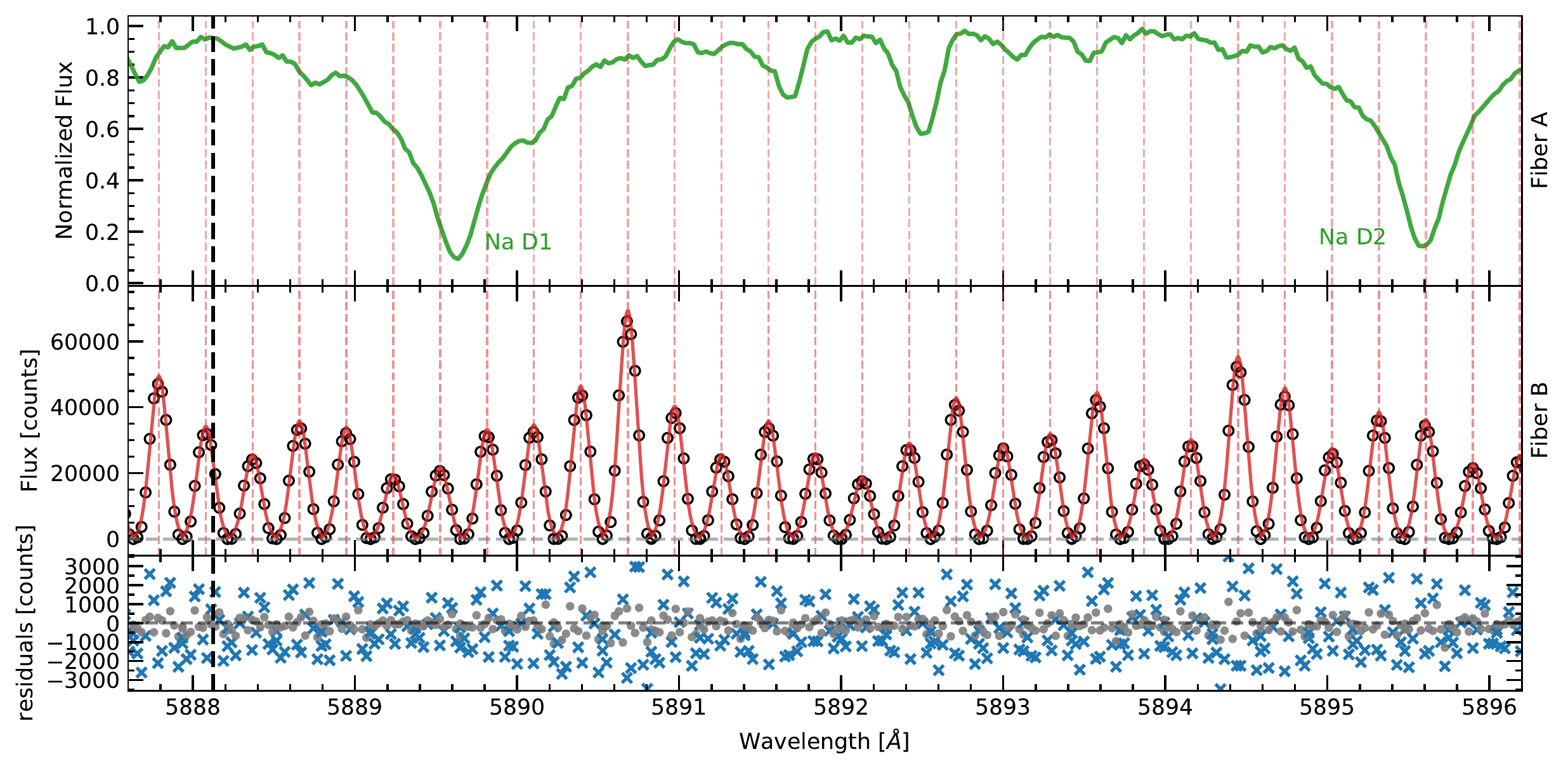}
    \caption{A part of spectrum of the $\tau$-Ceti centered around the double Sodium lines observed by HRS on 2.16m telescope with the simultaneous wavelength calibration by the 25GHz astro-comb. The top panel shows the normalized spectrum of $\tau$-Ceti fed in the science fiber with the wavelength range from about 5887.5\AA  \ to 5896.2\AA. The middle panel displays the LFC signal fed in the reference fiber. The red curves are Gaussian profiles fitted to each comb line individually. The red dashed vertical lines are their line center marked as the $\mu$  positions of Gaussians. The fitting residuals are shown in the bottom panel. The blue dots are residuals of Gaussian fit and the gray ones are residuals of Double Gaussian fit(Sect.\ref{sec3}). The black dashed line is the edge of a CCD mask at 2048(4 $\times$512) pixel position.}
    \label{Fig1}
\end{figure*}

Ideally, for a well-designed astro-comb system, each individual comb mode(line) should be well resolved at the giving line spacing and should be unresolved at the scale of the FWHM(no intrinsic structures).
With respect to the HRS spectrography for R$\sim$50,000 at 5500\AA, the 25GHz line spacing equals to 2.29 times of FWHM. This can be derived based on the equation below:
\begin{equation}
    |df| = \frac{c}{\lambda^2 } |d\lambda|
	\label{eq1}
\end{equation}
where c is the speed of light.
Considering that an infinitely narrow mode of LFC passes through the spectrograph, the instrumental response of the spectrograph will be the instrumental profile(IP). Note that the LFC's mode is much narrower than the resolution of our spectrograph, the comb line represents the IP and can be used as a tool to model the line profile and study the asymmetries and shape variation as a function of wavelength, pixel position and signal intensity etc.
In general, the comb signal can be described as:
\begin{equation}
    C(x) = \sum_{i=1}^N [ E_i \int_{i-n}^{i+m}IP(x)dx ]
	\label{eq2}
\end{equation}
Where $x$ is the pixel position in the spectral direction on each echelle order and $i$ is the line center for each comb line while $m$ and $n$ are the ends position  for each individual comb line. $E$ is the envelope function. $N$ is the total number of the comb lines in each echelle order.
For a 18GHz line spacing comb on R=110,000 spectragraph, $N$ is about 275$\pm$5 while for a 25GHz comb on R=50,000, $N$ is around 345$\pm$5 depending on each order's wavelength range and average SNR.
Thus, the LFC can be use as a powerful tool to characterize the line profile variation across the bandpass, to trace the instrumental shift at long-term and to probe the detector effects that induce errors to the RV measurements.
The significance of studying the line profile consist of:
1). Although the comb modes are filtered to eliminate side modes, the superfluous modes are not completely removed\citep{2010RScI...81f3105Q, 2012OExpr..20.6631Y} and the mode remnants may have influences on the line profile that can provide asymmetries and potential shift to the line center. This can be reflected by studying the comb line profile.
2). The signal of LFC undergoes two non-linear process during the optical path. One is the frequency doubling, the another is the spectra broadening. They will have impacts on the remnant side modes and may induce unwanted shift to the RV measurements. 
3). The errors from detector effects such as quantization uncertainty\citep{2014IAUS..293..407Z} and the influence of the CCD stitching effects can be studied in details by scanning all the pixels with the LFC lines.

This paper is structured as follows: In Section \ref{sec2} we explain the instrumental setup of the LFC and describe the data reduction methods. In Section \ref{sec3} we demonstrate the new model of line profile, estimate the precision derived from various models and analyze the parameters. In Section \ref{sec4} we discuss our results, and we present the conclusions and summaries in Section \ref{sec5}.


\section{Instrumental setup and Data reduction}
\label{sec2}

Our astro-comb calibration system is based on a commercial mode-locked Yb-fiber Menlosystem laser frequency comb\citep{2012Natur.485..611W} with the mode repetition rate of 250MHz which is locked to a rubidium atomic clock. Three Fabry-Perot cavities are employed as filters to suppress the unwanted intermediate modes and increase the line spacing to 25GHz which can be well resolved by the HRS spectrograph at R$\sim$50,000. Then we use a high power amplifier to reduce the affects due to low pulse energies. After the amplification, a second harmonic generator(SHG) is adopted to double the frequency of the light and converts the center of  wavelength coverage to the optical spectral region. 
We then use photonic crystal fibers(PCFs) to broaden the spectrum and obtain sufficient wavelength coverage.
The LFC signal is then coupled to a multimode fiber.
A fiber scrambler\citep{2016SPIE.9908E..7EY} is then implemented to increase the occupancy of the spatial modes of the light from the multimode fiber. By averaging a large number of modes, the scrambler can reduce the sensitivity of the light injection. Finally, the light from LFC is coupled to the HRS calibration fiber and illuminates the CCDs.

We installed the laser frequency comb in HRS spectrograph on 2.16m telescope at Xinglong observation in early 2016 with the efforts of the teams from Nanjing Institute of Astronomical Optics \& Technology (NIAOT), National Astronomical Observatories of China(NAOC) and Menlosystem. Several measurement campaigns has been carried out during the past two years. We highlight some typical acquisitions obtained from the HRS-LFC system and report the data analysis. In January 2017, we measured the performance of the wavelength calibration for this system. During this campaign, we obtained more than 60 consecutive exposures in one observing night with the light of LFC in the reference fiber(diameter=2.4") by using the typical exposure time of 20s and read-out time for the CCD of 40s.
The CCD is a E2V 4k$\times$4k 12$\mu$ scientific chip working at -106$^{\circ}$C cooling with Liquid nitrogen. The tempreature variation in short-term is $\pm$0.05$^{\circ}$C and $\pm$0.34$^{\circ}$C in a week\citep{2016PASP..128k5005F}.
A python and iraf based pipeline is used to reduce the data from the 2D raw fits to the flat-fielded, background-subtracted and extracted 1D spectrum. The default wavelength frame is measured by a ThAr lamp. The wavelength of comb line $i$ can also be computed independently by the setup of the LFC with the equation $\lambda_i = c/f_i=c/(f_0+n \cdot f_{rep})$ for giving $f_0$=9.52GHz, $f_{rep}$=25GHz and set the range of $n$ in [16000, 25000].
In each acquisition obtained in this campaign, the wavelength coverage of LFC is about 2300 \AA  (from 5085\AA \ to 7382\AA). There are 35 echelle orders(25th to 60th) can be used with high enough signal to noise ratio(SNR). The averaged SNR in the whole wavelength coverage is 234.9 with the two ends lower than 50. For each echelle order $j$ we can estimate the wavelength calibration precision $\sigma_j$ by calculating the rms(root mean square) around the wavelength solution curve. 
We find that when j<25 and j>60, we have $\sigma_j$>3$\sigma_{total}$ where $\sigma_{total}$ is the mean of $\sigma_j$ with 25<j<60.
Thus, with the limitation of both SNR and $\sigma_j$, we set the available comb line range between echelle order 25th and 60th, which is corresponding to [5085\AA ,\ 7382\AA].

In October 2017, we observed the $\tau$-Ceti(HD10700) in simultaneous calibration mode with the starlight in the science fiber and LFC light in the reference fiber. A set of neutral density(ND) filters were employed to control the signal intensity of LFC in the aim of exposing appropriately with the star. For a 200s exposure and ND=3dB setup, Fig \ref{Fig1} shows a part of the reduced spectrum in the echelle order 43th for two channels simultaneously. The mean FWHM(Full width at half maximum) of this order is 0.125\AA.  
We fit each individual comb line with Gaussian model. The red profiles in the middle panel of Fig \ref{Fig1} represent the Gaussian fitting results. Their residuals are plotted in the bottom panel with blue dots. 
The rms of residuals of Gaussian model in the wavelength range of [5888\AA, 5896\AA] is 1348.7 counts $e^-$ while for the double Gaussian model it is 376.9 counts $e^-$. We can see from Fig \ref{Fig1} that each comb line occupied about $\sim$14 pixels in this spectrum range which represents 0.288\AA.

We noticed that the amplitude of each individual comb line varies along the spectral direction. In Fig \ref{Fig1}, the maximum flux is 66112 counts $e^-$ and the minimum is 17579 counts $e^-$. Currently the reason of this variation is not fully understood\citep{2017sgvi.confE..30M}. It may come from the non-linear process during the optical pass or the affects from the fibers  or the combination of them. One of our aim is to study whether this variation gives influence to the wavelength calibration precision which is discussed in the following Section \ref{sec3}. We also notice that the Gaussian fitting residuals of each pixel is changing with its intensity. It can be seen from the bottom panel of Fig \ref{Fig1}. For a pixel near the peak of one comb line with the flux of 66112 counts $e^-$, the Gaussian fitting residuals of this pixel is 2989 counts $e^-$. For a pixel located near the gap between two adjacent lines, its residuals are close to zero.  As can be seen from Fig \ref{Fig1}, the double Gaussian model(gray dots) can improve this effect. In the following sections, we demonstrate the several models of comb line profile and discuss the correlation between the models and wavelength calibration precision.
%



\section{Characterizing the line profile with various models}
\label{sec3}

Fitting an appropriate model of comb lines can help develop a better understanding of the RV uncertainties induced by the line profile variation.
A widely used model of the the IP is the sum of several Gaussians\citep{1995PASP..107..966V}, which parameterizes a central Gaussian plus several satellite Gaussians.
The main component is chosen as Gaussian because the instrumental profile at first order follows a Gaussian profile. The satellite Gaussians as components are used to generate asymmetries to the whole profile.

\subsection{Optimized double Gaussian Model}

Recalling that in a typical exposure acquisition, there are about 12$\sim$14 pixels for each individual comb line. In this case, we mainly adopt two Gaussians to fit the line profile to achieve a high enough degree of freedom(DOF).
We use the Levenberg-Marquardt algorithm for fitting to find local minima.
In particular, we set a constraint to the parameters of double Gaussians to avoid unwanted fitting results. 
These situations occur mainly to the low SNR comb lines where the line wings may suffer from the remnants of side modes and the other noises. 
When constraining the parameters of double Gaussians to an optimized range, they can properly represents the intrinsic components of line profile and give out a better goodness of fit.

\begin{figure}
	\includegraphics[width=\columnwidth]{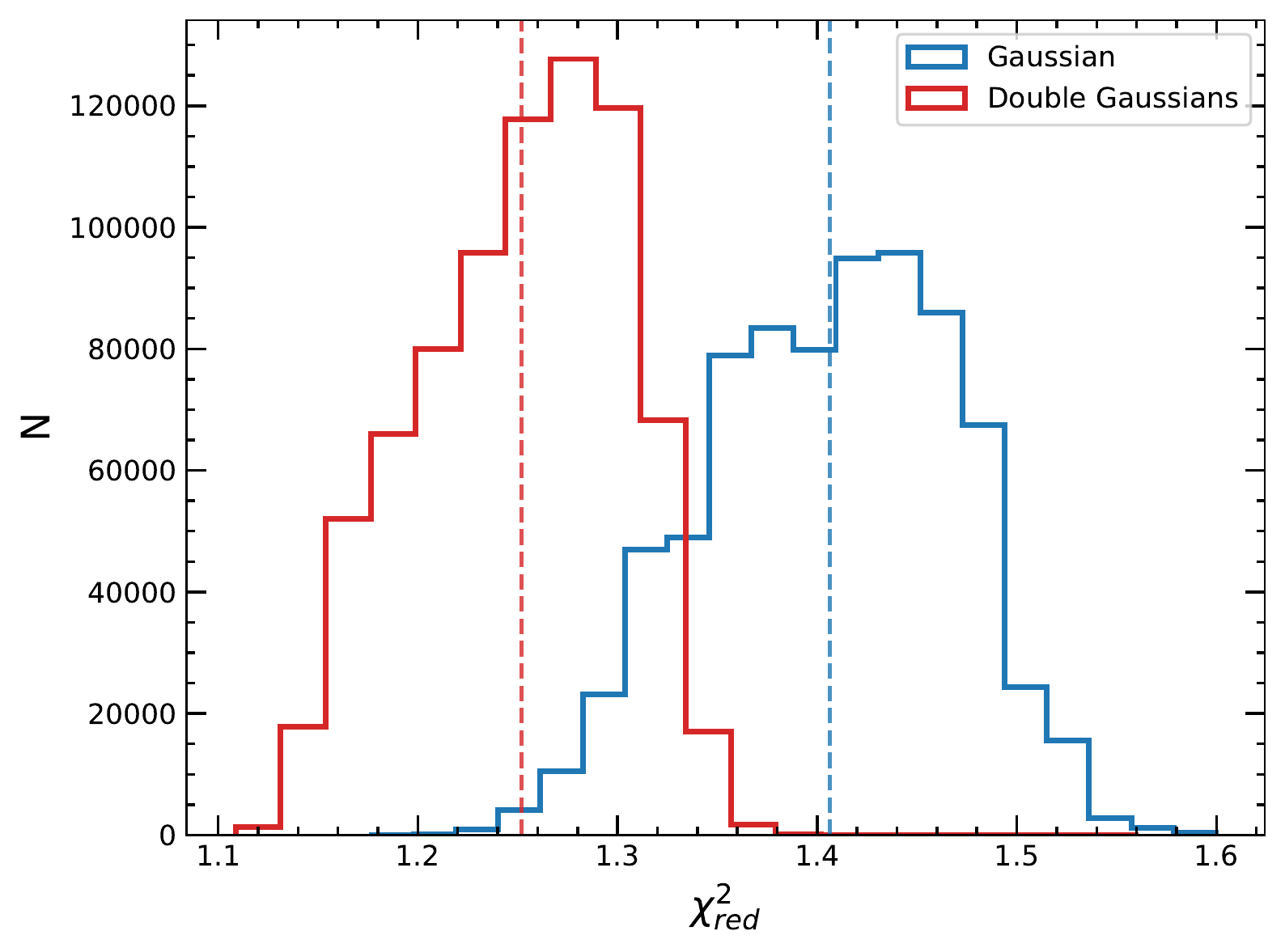}
    \caption{The distribution of the reduced $\chi^2$ for Gaussian model and the CDG model for comparison. The red dashed vertical line is the mean of CDG model, and the blue dashed vertical line is the mean for Gaussian model. The samples are all the 765337 comb lines in consecutive 63 exposure acquisitions.}
    \label{Fig2}
\end{figure}

Considering that $x$ is the pixel or wavelength unit along the spectrum's main dispersion direction, and $G_1$ and $G_2$ represent the two Gaussians function: $G_1(x, A_1, \mu_1, \sigma^2_1)=A_1 exp[-(x-\mu_1)^2 / 2\sigma_1^2]$ and $G_2(x, A_2, \mu_2, \sigma^2_2)=A_2 exp[-(x-\mu_2)^2 / 2\sigma_2^2]$. We assume that the line profile reaches the maximum when x=$x_m$. This can be described as:
\begin{equation}
  \frac{\partial G_1 + \partial G_2}{\partial x}\bigg|_{x = x_m} = 0
\label{eq3}
\end{equation}
With this equation, we can simply derived that:
\begin{equation}
x_m = \frac{\frac{\mu_1 G_1}{\sigma_1^2} + \frac{\mu_2 G_2}{\sigma_2^2}}{\frac{G_1}{\sigma_1^2} + \frac{G_2}{\sigma_2^2}}= \frac{\sigma_2^2 G_1 \cdot \mu_1 + \sigma_1^2 G_2 \cdot \mu_2}{\sigma_2^2 G_1 + \sigma_1^2 G_2}
\label{eq4}
\end{equation}
In this way, if we define a parameter $W$ as the weights:
\begin{equation}
W = \frac{\sigma_1^2 G_2}{\sigma_2^2 G_1 + \sigma_1^2 G_2}
\label{eq5}
\end{equation}
where 0<W<1. Then Equation \ref{eq4} can be written as:
\begin{equation}
x_m = \mu_1 - W \cdot (\mu_1-\mu_2)
\label{eq6}
\end{equation}
When $W=0$ we can get $x_m = \mu_1$ and when $W=1$ we have $x_m = \mu_2$. Equation \ref{eq6} indicates that the maximum position(mode) lies in the range of [$\mu_1$, $\mu_2$]. The distance between $\mu_1$ and $\mu_2$(marked as $\left| \mu_1 - \mu_2 \right|$) can be used to set constraints to the two Gaussians components. Considering a free $\left| \mu_1 - \mu_2 \right|$ situation, when there is a 'jump' data point near one of the line wing due to certain noises or side mode remnants, the satellite Gaussian profile may arrange its peak around this jump area to give more weights and to find minimum chi-square for the whole profile. This could generate unwanted fitting results such as 'double-peaks' shape which are obviously not the real intrinsic profile.
We found this phenomenon in the comb exposure acquisitions especially at low SNR level. Typically, taking the 45th echelle order for example($\sim$ 356 lines), we compared the 45th orders from the exposure acquisitions of different SNR. When the mean SNR is less than 60 the 'double-peak' lines are 1.12\%. When the mean SNR<50 the ratio grows to 1.96\%.
To avoid this situation, one of the conservative estimation for the constraint is to use:
$\left| \mu_1 - \mu_2 \right| < FWHM$ or $\left| \mu_{1,2} - \mu \right| < \frac{1}{2} FWHM$ where $\mu_1$ and $\mu_2$ are centers of the two components Gaussians $G_1$ and $G_2$ and $\mu$ is the center of single Gaussian fit. To express the FWHM, we use the relations of $A \cdot exp[-(x-\mu)^2 / 2\sigma^2] = (1/2) \cdot G_{max}$ and $G_{max} = G(\mu)$ to derive that $FWHM = 2\sqrt{-2 \cdot ln(1/2)} \sigma \approx 2.355 \sigma$. Then the constraint correlation can be written as:
\begin{equation}
   \mu - 1.177 \sigma < \mu_{1,2} < \mu + 1.177 \sigma
\label{eq7}
\end{equation}
Here $\mu$ and $\sigma$ are calculated from single Gaussian fit.
Considering a low SNR line with certain asymmetric profile, the $\mu$ and $\sigma$ from Gaussian fit may already induce errors compared to the real line center and line width. Thus, we use the first raw moment and second central moment instead of $\mu$ and $\sigma$. For the $i$th pixel with flux $x_i$, the first raw moment equals to center of gravity(CG here after) which is $CG = E[x_i]$. The second raw moment is the variance of the $x_i$ distribution which can be written as: $V = \sqrt{E[x_i - CG]^2} = K^{-1} \cdot \sigma$. With regards to our situation that each comb line occupied about 12$\sim$14 pixels, we computed the coefficient $K=1.107\pm0.000874$.
In this way, the constraint conditions can be described as:
\begin{equation}
   \left| \mu_{1,2} - CG \right| < \sqrt{2 ln2} K\cdot V
\label{eq8}
\end{equation}
where the first raw moment and second central moment can be calculated directly from the extracted 1D spectra without any fitting process and its errors.
For the example we mentioned above, by using this constraint to the 45th ehcelle order for double Gaussians fitting, we find no unwanted fitting situation(e.g. double-peak shape) in the whole process.
We apply this constrained double Gaussians fitting method to all the comb lines in our acquisitions(765337 lines in consecutive 63 exposures) and compared it to other fitting models.

\begin{figure}
	\includegraphics[width=\columnwidth]{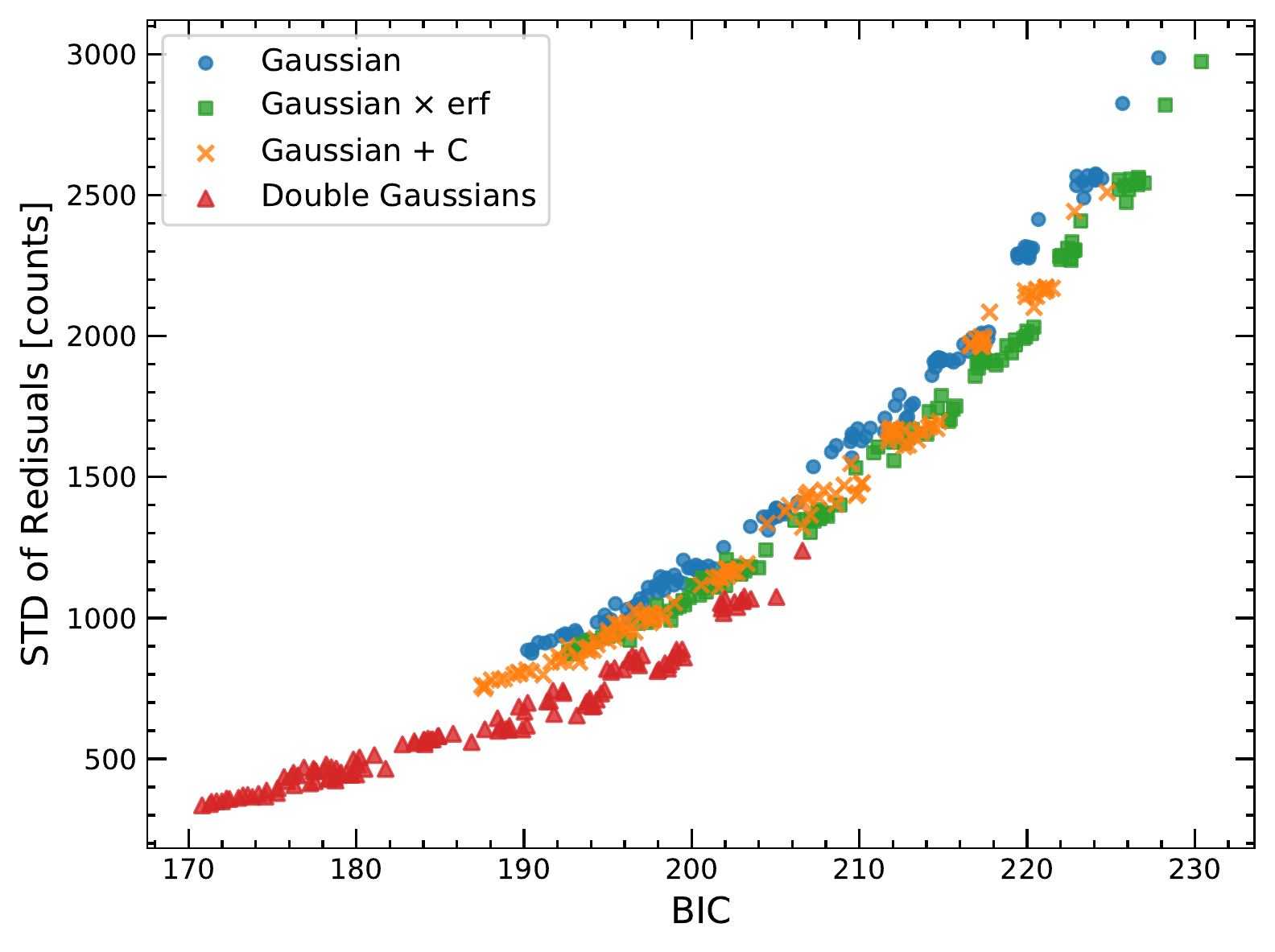}
    \caption{The comparison of various models. We use two typical ehcelle orders(32th, 33th) to measure BIC and the fitting residuals for example. The other orders have similar correlation and range like this. Each point represent the mean value of the all the comb lines in the orders. The y-axis is the standard deviation of the fitting residuals, which describe how scattered the residuals are.}
    \label{Fig3}
\end{figure}

To evaluate the goodness of the fit, one of the direct approaches is to calculate the reduced $\chi^2$ for each individual line. Figure \ref{Fig2} shows the comparison of the single Gaussian and the constrained double Gaussians(CDG hereafter) fitting. The mean reduced $\chi^2$ of the Gaussian model is $<\chi^2_{red, G} >= 1.406$ with the standard deviation of 0.0611. For the CDG model, we have  $\chi^2_{red, CDG} = 1.252 \pm 0.0505$. The CDG model has $\chi^2_{red}$ more closer to 1 in most cases. However, we should be careful about adding more parameters to increase the likelihood by considering the situation of  overfitting. For a more stringent study of the model selection, we attempt to use another advanced indicator in the next subsection with the aim to analyze which model is preferred.

\begin{figure}
	\includegraphics[width=\columnwidth]{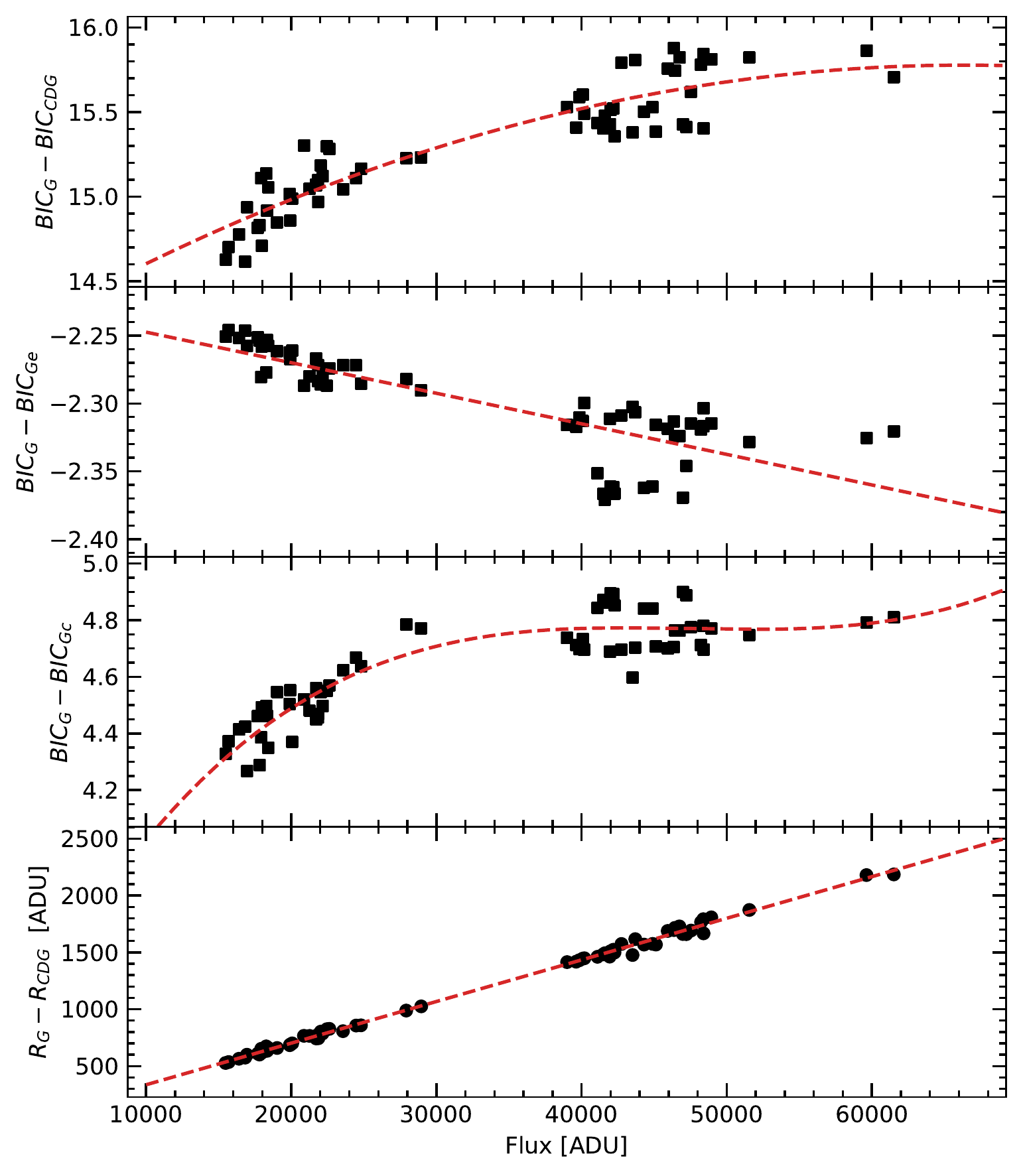}
    \caption{The correlation between $\Delta BIC$ and the averaged line intensity grouped by each exposure acquisition. In the top panel, $BIC_{CDG}$ stands for the BIC values of constrained double Gaussian model and $\Delta BIC$ is the difference between Gaussian and CDG model. The second panel and third panel show the $\Delta BIC$ of Gaussian$\times$error function model and Gaussian+constant model respectively. The bottom panel is the difference of fitting residuals for Gaussian against CDG model. The red dashed lines are polynomial fitting curves.}
    \label{Fig4}
\end{figure}

\subsection{BIC analysis of various models}

\begin{figure}
	\includegraphics[width=\columnwidth]{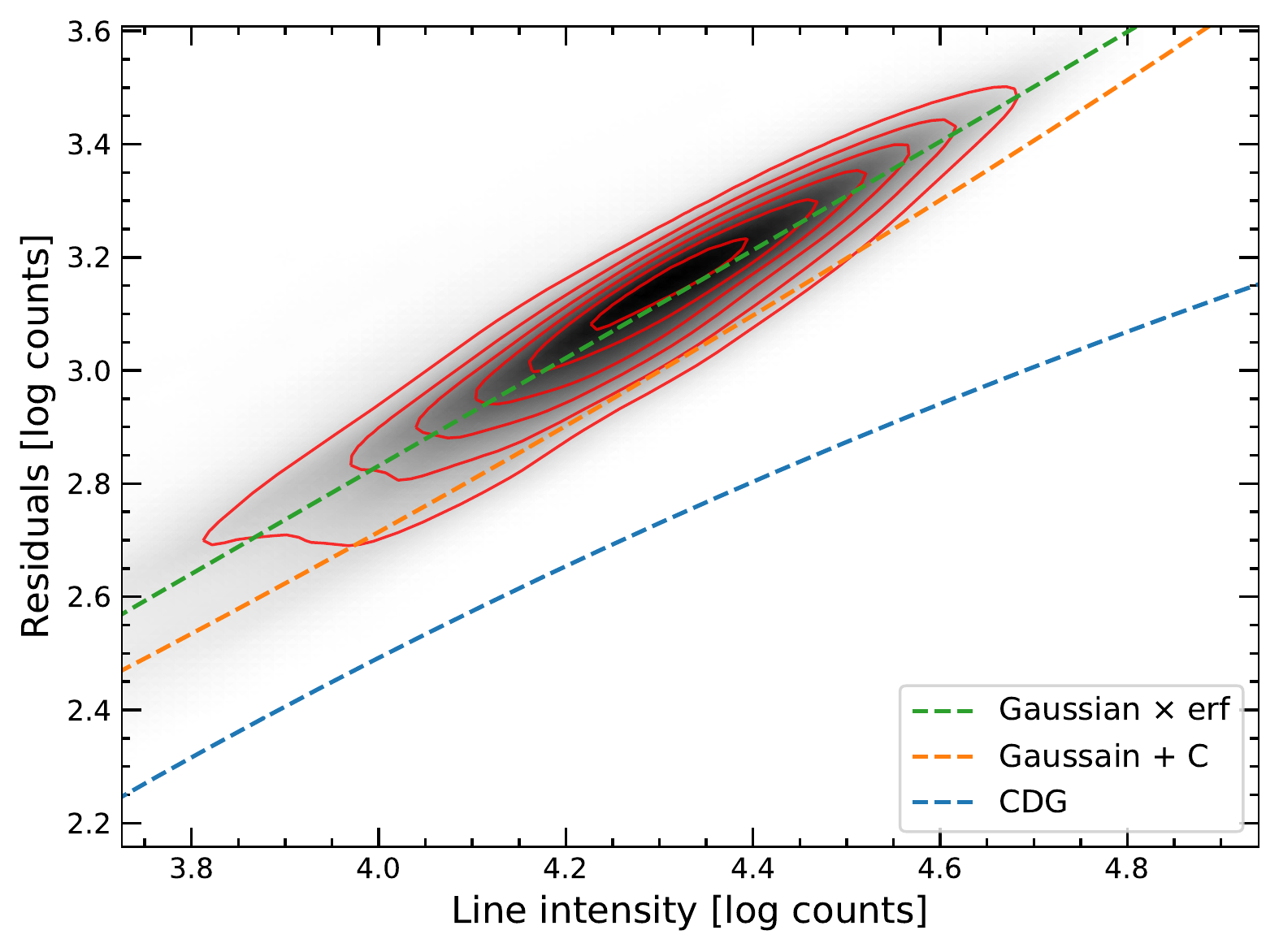}
    \caption{The comparison of different models with the fitting residuals and comb line amplitude. The transparent black dots are samples of Gaussian fit for all the available comb lines(11251) in a typical exposures. The density is described with contours marked as red lines. For the other models, the three dashed lines are fitting to their areas, without plotting the raw dots for a clear comparison. Both the x and y axis are in log scale.}
    \label{Fig5}
\end{figure}

In statistics, one of the pervasively used tools in candidate model selection is the Schwarz criterion\citep{schwarz1978} or Bayesian information criterion(BIC, hereafter)\citep{doi:10.1080/01621459.1995.10476592, 2007MNRAS.377L..74L} which is an asymptotic approximation applied to a form of the posterior probability in Bayesian statistics for determining candidate models.
The standard definition of BIC is as follow:
\begin{equation}
   BIC = -2ln[L(\hat{\theta}_k | x)] + k \cdot ln(n)
\label{eq9}
\end{equation}
where $x$ is the observed data, $n$ is the sample size, e.g. the number of data points in x, $k$ is the number of parameters in the model and $\theta_k$ stands for the set of all parameters. The components in the k-dimensional parametric vectors are functionally independent. The $L(\theta_k | x)$ represents the likelihood corresponding to the density function $p(x|\theta_k)$. The $\hat{\theta}_k$ denotes the estimated parameter values obtained by maximizing the likelihood function.
For the line shape analysis, we assume that the errors of the different models are independent and identically distributed following a normal distribution. 
We also suppose that the derivative of the likelihood in log scale as the boundary condition with respect to the variance is zero\citep{nla.cat-vn2888327}. In this case, equation \ref{eq9} can be derived as:
\begin{equation}
   BIC = n \cdot ln \left[ \frac{\sum_{i=1}^n (Resid_i)^2}{n} \right] + k \cdot ln(n)
\label{eq10}
\end{equation}
Here $Resid_i$ denotes the fitting residuals of position $i$. In this way, we can calculate the BIC value for each comb line and for different models such as n=14 and k=3 as Gaussian profile.
Comparing various models with the Bayesian information criterion can simply refer to calculate the BIC value for each model. The model with the lowest BIC value is considered the best one.

Figure \ref{Fig3} shows the BIC comparison for four models: Gaussian, constrained double Gaussians, Gaussian multiplied by an error function and Gaussian plus a constant. The error function is set with the aim to help generate skewness components to find the best fit especially for the comb lines at low SNR level. A constant as free parameter is added to a Gaussian for considering the contributions of the remnants of background. 
For CDG model, it has the lowest BIC value's range which is from 170.8 to 206.6. With regards to the other three models, the BIC's range are similar which is about [187.4, 230.38]. The 'Gaussian+C' model is relatively $\sim$7\% better for the BIC's range. 
Recalling from Fig.\ref{Fig1} bottom panel, the residuals have different distributions when the line intensity varies. We measure the standard deviation of the fitting residuals and grouped by each order as it is shown in y-axis of Fig.\ref{Fig3}. 
We find that the "flux-flat" orders tends to have smaller BICs. While the orders with stronger variation of line intensity may have larger BICs. For the CDG model, this correlation follows a 2th-order polynomial:
\begin{equation}
   STD_{Res} = 1.206 \times 10^4 + 143.5BIC + 0.439BIC^2
\label{eq11}
\end{equation}
When we study the target comb lines in a certain wavelength range or pixel position range, we are able to simply obtain the $STD_{Res}$ by the fitting code and then we use the equation \ref{eq11} to estimate the BIC values and check the goodness of fit.

\begin{figure}
	\includegraphics[width=\columnwidth]{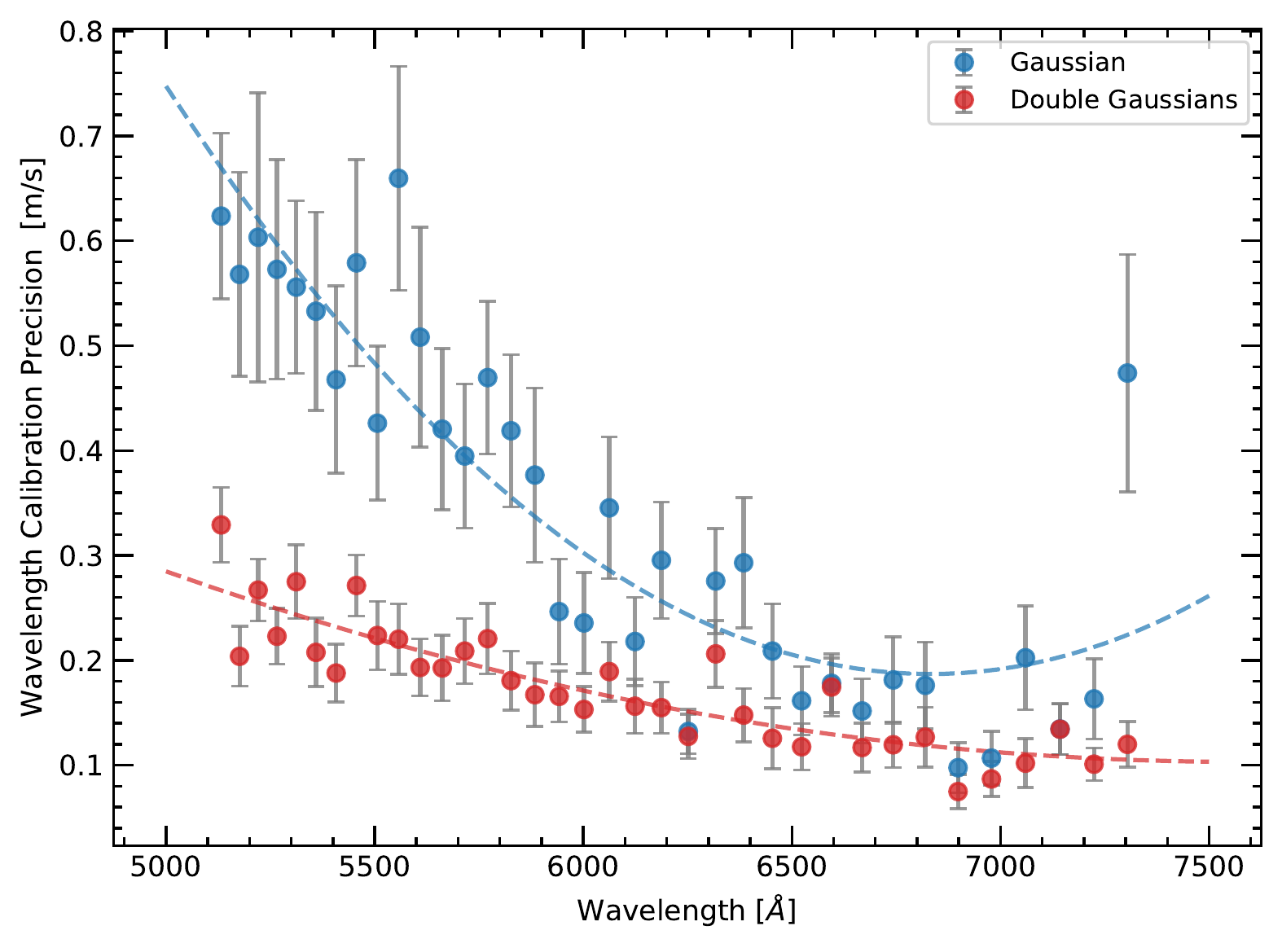}
    \caption{The wavelength calibration precision of Gaussian model and CDG model grouped by each order. The data is from 63 consecutive exposures. The blue and red dashed line are polynomial fitting curves for Gaussian and CDG respectively.}
    \label{Fig6}
\end{figure}

We also calculate the $\Delta BIC$ for the comparison, which is the difference of BIC values between a particular model and the target model. The $\Delta BIC$ can be used as an argument against the other models. 
Generally, if the value of $\Delta BIC$ is between 2 and 5, we can say the evidence against the target model is positive.
If $\Delta BIC$ is between 5 and 10, it can be believed that the evidence for the best model compared to the weaker model is strong. If $\Delta BIC>10$, it means the evidence for the best model against the alternate models is very strong\citep{2007MNRAS.377L..74L}.
We choose the Gaussian as the target model, and use $\Delta BIC$ to evaluate the other models compared to Gaussian.
Figure \ref{Fig4} depicts the $\Delta BIC$ changing as a function of the line intensity. 
For each individual comb line $i$, assuming it has $\sim$14 data points, we calculate its BIC values of Gaussain model, Gaussian$\times$error function model(Ge, hereafter), Gaussian+constant model(Gc, hereafter) and of CDG model by computing the difference of them. Then they are grouped by each exposure acquisition which is displayed as the black dots in the top panel of Figure \ref{Fig4}. During the measurements, we used neutral density filters (0.2dB to 4dB) to control the average light intensity of the exposures. 
For the CDG model, we find that the $\Delta BIC$ increases when the averaged flux of lines' peak is growing. For the exposure with lowest mean flux, it has the $\Delta BIC = 14.61$, while for the acquisitions with mean flux of peak greater than $6 \times 10^4$, the $\Delta BIC$ increased to $>15.7$. The whole trend in our measurement range follows a polynomial as:
\begin{equation}
   \Delta BIC = 14.15 + 4.91 \times 10^{-5} F - 3.7 \times 10^{-10} F^2
\label{eq12}
\end{equation}
Where $F$ denotes the flux as the averaged comb line intensity for each exposure acquisition.
This evidence means that the CDG model is significantly better for fitting comb lines in all the flux range and especially tends to have better fitting results for the stronger lines.
Equation \ref{eq12} can be used to estimate the $\Delta BIC$ in various flux levels for given exposures in our following observations. 
For the Ge model, we find it works worse than Gaussian due to the negative $\Delta BIC$ in the range of (-2.38, -2.2). With regard to the Gc model as its $\Delta BIC$ in (4.2, 5), it shows a positive evidence against the Gaussian model.
On the other hand, we already notice that the strong lines has larger residuals(e.g. Fig. \ref{Fig1}, bottom panel) for Gaussian profile fitting. The CDG can effectively reduce the fitting residuals for the comb lines in high flux level.
In the bottom panel of Fig.\ref{Fig4}, averagely speaking, the difference between the fitting residuals of Gaussian and CDG grows up with the increase of peak flux, which is roughly along a linear function as $\Delta Resi = -30.5 + 0.0366 \times Flux_p$. This relationship is corresponding to equation \ref{eq12} which indicates that the CDG model fit the line profile better especially for strong lines.

\begin{figure}
	\includegraphics[width=\columnwidth]{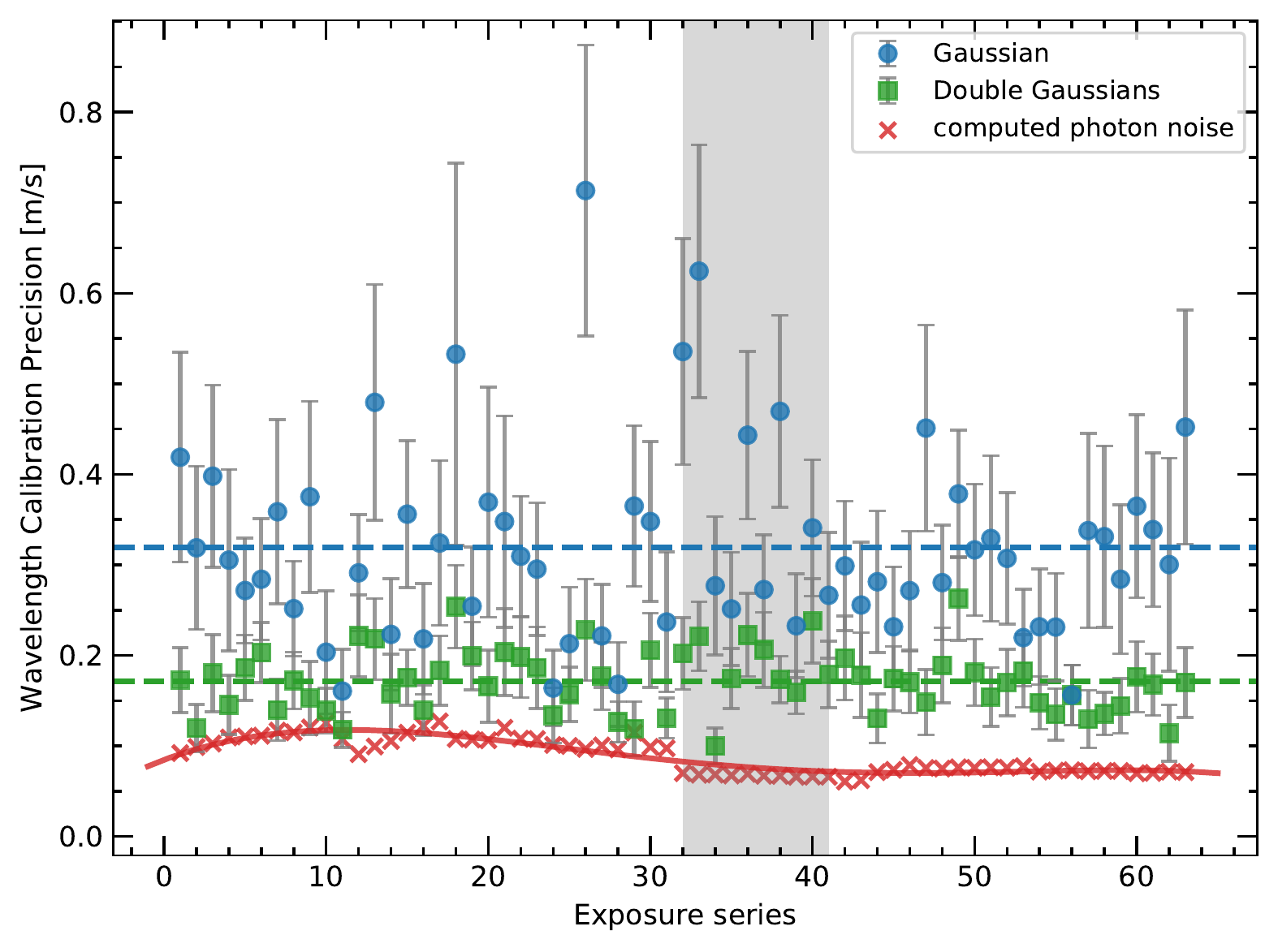}
    \caption{The comparison of photon noise limit and the precision derived from the two models. The blue and green dashed lines are the means of the distribution of Gaussian and CDG model. The red line is the polynomial fit of photon noise. The gray shadow area marks the 10 consecutive exposures with relative stable photon noise values.}
    \label{Fig7}
\end{figure}

\begin{table*} 
	\centering
	\caption{The detail information of the 10 chosen exposures(marked as gray shadow) in Fig.\ref{Fig7}. The first column is the series number . The second and third columns are wavelength calibration precision or uncertainty of Gaussian and CDG model respectively.}
	\label{tab1}
	\begin{tabular}{ | c | c | c | c | c | c | c | } 
		\hline
		Series no. & Uncertainty(G) [m/s] & Uncertainty(CDG) [m/s] &Photon noise [m/s] & Quality factor  & FWHM [$\overset{\circ}{A}$] & SNR \\
		\hline\hline
        1	&    $ 0.53 \pm  0.12	$ &  $ 0.20 \pm  0.039  $ &  0.07311	& 60617.15	&   $ 0.12612  \pm 0.000148 $ 	& 217.44     \\
        2	&    $ 0.62 \pm  0.13	$ &  $ 0.22 \pm  0.038  $ &  0.07124	& 59872.55	&   $ 0.12614  \pm 0.000148 $ 	& 225.40     \\
        3	&    $ 0.27 \pm  0.07	$ &  $ 0.09 \pm  0.019  $ &  0.07147	& 59327.83	&   $ 0.12614  \pm 0.000148 $ 	& 226.45     \\
        4	&    $ 0.25 \pm  0.06	$ &  $ 0.17 \pm  0.033  $ &  0.07036	& 60406.28	&   $ 0.12607  \pm 0.000148 $ 	& 226.62     \\
        5	&    $ 0.44 \pm  0.09	$ &  $ 0.22 \pm  0.046  $ &  0.07213	& 60747.34	&   $ 0.12608  \pm 0.000148 $ 	& 219.84     \\
        6	&    $ 0.27 \pm  0.06	$ &  $ 0.20 \pm  0.041  $ &  0.07004	& 59345.02	&   $ 0.12617  \pm 0.000148 $ 	& 230.96     \\
        7	&    $ 0.46 \pm  0.10   $ &  $ 0.17 \pm  0.025  $ &  0.06990	& 60607.95	&   $ 0.12617  \pm 0.000161 $ 	& 227.36     \\
        8	&    $ 0.23 \pm  0.05	$ &  $ 0.15 \pm  0.023  $ &  0.06884	& 58243.03	&   $ 0.12612  \pm 0.000148 $ 	& 238.80     \\
        9	&    $ 0.34 \pm  0.07	$ &  $ 0.23 \pm  0.046  $ &  0.06900	& 60279.56	&   $ 0.12618  \pm 0.000148 $ 	& 231.34     \\
        10	&    $ 0.26 \pm  0.06	$ &  $ 0.17 \pm  0.036  $ &  0.06888	& 60038.58	&   $ 0.12610  \pm 0.000149 $ 	& 232.65     \\
		\hline
	\end{tabular}
\end{table*}

The relationship between the amplitude of comb line and the fitting residuals for various models is shown in Fig.\ref{Fig5}. We find that 35.5\% comb lines concentrate in the line amplitudes range of [4.2, 4.4] in log scale. The mean of residuals for Gaussian model is 1279.2 counts $e^-$, comparing with the CDG model as $550.5 \pm 251.1$ counts $e^-$. With regards to Gaussian$\times$error function model, we estimate its residuals as $Res_{Ge} = 1265.4 \pm 631.5$ counts $e^-$ which is almost the same with Gaussian fit. For the model of Gaussian+constant, we have $Res_{Gc} = 992.1 \pm 540.8$ counts $e^-$ that is about 22.4\% better than Gaussian fit. From the comparison, it is evident that CDG model is currently the best one. 
In the next step, we will discuss the results 
when we apply the CDG model to the observed comb data.

\subsection{Improving the RV precision by CDG model}

In the subsections above, we introduce the CDG model and demonstrate that it has a better goodness of fit than other models for comb line profile. Thus, an important question may arises that how the CDG model can improve the wavelength calibration precision. In this section, we discuss the performance when applying the CDG model to the comb data.

To be clarified, there are several definitions of RV precision referring to Doppler method which basically stem from two approaches: iodine cell technique\citep{1996PASP..108..500B} and cross correlation technique\citep{2007A&A...468.1115L}. For different time scales, they can be classified as single measurement precision and long-term radial velocity rms. To investigate the LFC as the wavelength calibrator, we adopt the method of \citet{2012Natur.485..611W}, in which the single measurement precision is calculated from the standard deviation around the wavelength solution curve for each echelle order. 
\begin{equation}
   \sigma_i = std \left[ \frac{3 \times 10^8 \times (\lambda_j - \hat{\lambda}_j)}{\lambda_j} \bigg|_{j=1,2...N}  \right]
\label{eq13}
\end{equation}
here for the $i$th order, $\lambda_j$ is the wavelength of the $j$th comb line and $\hat{\lambda}_j$ stands for its wavelength values from the wavelength calibration curve(5th order polynomial). The N is the total number of the comb lines in the $i$th order.

Figure \ref{Fig6} shows the wavelength dependence of the precision. We calculated the $\sigma_i$ for $i$ in [25, 60] and grouped by each order. For the Gaussian fit situation, there is a steep decrease trend from about 5100\AA to 6600\AA. This trend grows up near the red end of the coverage. With regards to the orders whose wavelength is less than 5500\AA, their mean precision is 0.56$m \ s^{-1}$ and maximum is 0.62$m \ s^{-1}$. It drops to the minimum of 0.097$m \ s^{-1}$ at 6898\AA. Then it increases to 4.74$\pm$1.13 $m \ s^{-1}$ at 7304\AA.
For the CDG model, the inclination is relatively weak and smooth. Approximately, it follows the function as:
\begin{equation}
\sigma_{CDG} = 0.676 + 0.000415 \times \lambda + 2.74 \times 10^{-8} \times \lambda^2
\label{eq14}
\end{equation}
As it is depicted in Fig.\ref{Fig6}, the precision of the red side is relatively better than the blue side. One of the reason is the varying line interval in different orders due to the comb's characteristic. Recalling equation \ref{eq1}, when $\Delta f$ is fixed to $f_{rep}$, we will have a larger $\Delta \lambda$ at greater $\lambda$. In this way, a larger distance between neighbour comb lines could reduce the influence from its conjoint neighbour comb lines and enable a better determination of each line's position from the fitting models. Thus, the results of wavelength calibration precision is better. However, on the blue side, the narrower interval distance makes the comb lines slightly suffer the signals from their neighbour lines or remnant of side modes between them. Besides, we also notice that there are several scatted points of Gaussian model near red end, which may be due to the low SNR in this area. The mean SNR in the order close to the red edge is 39.8. 
In a word, we find the precision derived by CDG model is generally better than traditional Gaussian model in the wavelength coverage shorter than $\sim$6500\AA. Near the blue end, the improvement of CDG model is about 56\%. 
In Fig.\ref{Fig6}, we also notice that there exists a outlying data point near the red edge of the comb wavelength coverage due to the low SNR level in the red end. If we only examine the comb lines without considering the red edge($\sim$7300\AA), the performance of the two models are similar at the wavelength longer than $\sim$6500\AA.

Figure \ref{Fig7} exhibits the precision grouped by each exposure acquisition. For the 63 consecutive exposures, the precision of CDG model is $\sigma_{CDG} = 0.17 \pm 0.034$ and for Gaussian model is $\sigma_G = 0.32 \pm 0.11$. During the whole test series, the improvement of CDG model is about 47\% compared with Gaussian model. We also investigate the photon noise limit of all these exposure acquisitions, and compare it to the two fitting models. An optimum photon-limited precision is claimed in \citet{2007MNRAS.380..839M}. We compute the $(S/N)_{max}=244$ for a typical acquisition. Then we estimate the precision by given:
\begin{equation}
\sigma^{opt} = 0.45 \cdot \left(\frac{500}{244} \right)  \left( \frac{1.5 \times 10^5}{R} \right)^{\frac{3}{2}} = 4.79 \ [cm \ s^{-1}]
\label{eq15}
\end{equation}
where R=50,000 for the current setup of HRS. We also compute the photon noise and quality factor for each exposure individually based on \citep{2001A&A...374..733B}. Considering a small enough RV change is detected from pixel $i$ with the intensity $I(i)$, the pixels with a larger gradient of flux tends to have more contributions to the RV precise, which can be expressed as:
\begin{equation}
\frac{\delta(i)}{c} = \frac{\Delta I(i)}{\lambda(i) \cdot \left[ \frac{\partial I(i)}{\partial \lambda(i)} \right]}
\label{eq16}
\end{equation}
where $\Delta I(i)$ is the change of intensity relative to a reference(epoch 0). By giving an optimum weight $W(i)=[\delta(i)/c]^{-2}$, one can derive the quality factor $Q=\sqrt{\Sigma W(i)} / \sqrt{\Sigma I_0(i)}$. Then the uncertainty of the RV change is given by:
\begin{equation}
\delta_{rv} = \frac{c}{Q \cdot \sqrt{N_{e^-}}}
\label{eq17}
\end{equation}
Here $N_{e^-}$ is the whole number of the photoelectrons gathered in the entire spectral range. By applying this algorithm, we calculate the photon noise limit(red dots in Fig.\ref{Fig7}) and the corresponding quality factor $Q$ for all the available exposure acquisitions as $\sigma_v = 0.088 \pm 0.019 m \ s^{-1}$ and $Q = 63018 \pm 2484$.
We pick out 10 successive and relative stable exposures with $\Delta \sigma_v = 1.3 \times 10^{-3} m \ s^{-1}$ and show the details in Table\ref{tab1}. From Fig.\ref{Fig7} and Table\ref{tab1}, we can see the average precision of CDG is about 46\% better than the Gaussian model but still above the photon noise limit.

\section{Discussion}
\label{sec4}

Using the asymmetric fitting models for line profile or calculating from the raw moment analysis, we find the comb line shapes are not perfectly symmetric. Approximately, the skewness of line shape can be seen as a function of line intensity and the line's position on the detector. The origins of the asymmetric component is still not clear. With the clues given by the analysis methods in this paper, it may source from the asymmetric instrumental profile, the CTI effect\citep{2005PASA...22..257T, 2012MNRAS.422..761M}, the imperfect light injection and the envelop-background problem of the current LFCs \citep{2017sgvi.confE..30M}, or a mixture of all the sources above which are estimated to generate about several tens of $cm \ s^{-1}$ uncertainties.
To understand the origins of line skewness can help push down the RV precision below 1 $m\ s^{-1}$ and keep stable for long-term scale.
In the next step, the CDG model is considered to be used as a tool to study the asymmetric properties of line profile by designing more experiments and observations.

There are several ongoing RV surveys that have achieved the RV precision from a few $m \ s^{-1}$ to the level less than 1 $m \ s^{-1}$, such as ELODIE\citep{1996A&AS..119..373B}, HARPS\citep{2002Msngr.110....9P, 2003Msngr.114...20M} and HARPS-N\citep{2012SPIE.8446E..1VC}. The up-coming instruments with the main goals to search for terrestrial planets at several tens of $cm \ s^{-1}$ precision includes ESPRESSO(Echelle SPectrograph for Rocky Exoplanets and Stable Spectroscopic Observations)\citep{2014AN....335....8P}, HARPS3\citep{2016SPIE.9908E..6FT}, EXPRES(The EXtreme PREcision Spectrograph)\citep{2016SPIE.9908E..6TJ} and CARMENES\citep{2014SPIE.9147E..1FQ}.
For the RV precision level below 1 $m \ s^{-1}$, some instrumental effects and uncertainties must be carefully studied. 
With the application of LFC and the CDG model, it is possible to characterize the spectrometer drift, estimate the PSF across the instrument bandpass and investigate the detector imperfections.
In the current test, when we apply CDG model analysis into the pipeline for routine observations, the computational cost of time is about extra 10 to 20 seconds for each typical exposure which contains about 12,000 comb lines. 

Moreover, the precise radial velocity measurements with LFC are significant for the science of the current and future space missions.
With the high precision RV follow-up observation with astro-comb, we can enhance the productivity of the transit missions such as TESS\citep{2014SPIE.9143E..20R}, CHEOPS\citep{2017SPIE10563E..1LC} and  PLATO\citep{2016AN....337..961R}.

\section{Conclusions and Summary}
\label{sec5}
We reported the successful observation and measurement with a 25GHz LFC system on HRS spectrograph in Xinglong observatory. In the measurement campaigns, we obtained more than 60 consecutive exposure acquisitions during one observing night. More than 2000\AA wavelength range(from 5085\AA to 7382\AA) is covered by comb lines in each exposure acquisition. With the statistic investigation about multitudes of comb lines as samples, we develop a novel model to fit the line profile.
The Double Gaussians model with a constraint can effectively give out a better goodness of fit even at low SNR level.
The constraint is given as $\left|\mu_{1,2} - \mu \right| <\sqrt{2ln2}\sigma$ where $\mu$ and $\sigma$ are from traditional Gaussian fit. To evaluate and compare the goodness of fit among various models, we introduce Bayesian information criterion to test the data. After comparison, it is obvious that the CDG model is significantly better than any other models. We apply the CDG model to the obtained comb data to testify the improvement of RV precision by CDG model. The results shows that the CDG model can give better wavelength calibration precision about 20 $cm \ s^{-1}$.
In the next step, we also consider to use the LFC and CDG model to characterize the line shape variation across the detector.

\section*{Acknowledgements}

We would like to thank the referee for the helpful comments that have improved the paper. We are grateful to the team in Xinglong observatory for their work to provide the data.
We also thank the team of NIAOT(Nanjing Institute of Astronomical Optics and Technology) for the setup of LFC and technical support.
This work is supported by the National Science Foundation of China(NSFC) under the grant numbers 11703052, 11390371 and 11303042.





\bibliographystyle{mnras}
\bibliography{bib} 









\bsp	
\label{lastpage}
\end{document}